\documentclass[12pt]{article}

\usepackage[dvips]{graphicx}
\begin{document}

%
%

\title{Enhanced Wavelet Analysis of Solar Magnetic Activity with Comparison to Global Temperature and the Central England Temperature Record}
%

%
%


\author{Robert W. Johnson\footnote{Alphawave Research, Atlanta, Georgia, USA. (robjohnson@alphawaveresearch.com)}}

\date{\today}
\date{February 25, 2009}

\maketitle



%
%
%

%
%


\begin{abstract}
The continuous wavelet transform may be enhanced by deconvolution with the wavelet response function.  After correcting for the cone-of-influence, the power spectral density of the solar magnetic record as given by the derectified yearly sunspot number is calculated, revealing a spectrum of odd harmonics of the fundamental Hale cycle, and the integrated instant power is compared to a reconstruction of global temperature in a normalized scatter plot displaying a positive correlation after the turn of the twentieth century.  Comparison of the spectrum with that obtained from the Central England Temperature record suggests that some features are shared while others are not, and the scatter plot again indicates a possible correlation.
\end{abstract}


%
%

%


%
%

\section{Introduction}
The continuous wavelet transform provides an interesting tool with which to explore the harmonic analysis of a signal. (See \cite{kaiser:1994} and \cite{torrence98practical} for friendly and practical guides.)  With a renormalization of the magnitude, its square returns the power spectral density estimated from the signal, and with the renormalization of transform coefficients outside the cone-of-admissibility, one may correct for the decrease in amplitude associated with the truncation of the wavelet by the edges of the known data.  Defining the wavelet response function by the power spectral density of sinusoids allows one to deconvolve the spectrum, giving a best estimate of the instantaneous wavelet power.

Recent examples of the use of wavelets in the analysis of sunspots are given by~\cite{ans-c49,cjaa:35391,msait:761026}.  Using the corrected and enhanced wavelet transforms, we conduct a spectral analysis of the solar magnetic cycles present in the record of sunspot numbers for the years 1700---2006.  The enhanced power spectral density reveals the time variation of the cycle periods, as well as that of their magnitudes.  The integrated instant power is compared to a record of global temperature as given by the glacial reconstruction of~\cite{Oerlemans:2005}, first on absolute axes and then as a normalized scatter plot, and also compared to the Central England Temperature (CET) record.  The enhanced spectra for the solar activity and the CET display similar features at the odd harmonics of the Hale cycle; however, such similarity may be only coincidental, and whether any true correlation exists between the magnetic cycles and global climate is still a matter of debate.

\section{Continuous Transform Using Truncated Morlet Wavelets}
For this analysis, we use the complex wavelet introduced by~\cite{Morlet:1984}, \begin{equation}
\Psi(s,t,t') \equiv \Psi_{s,t}(t') \equiv C_s h_{s,t}(t') \Phi_{s,t}(t') = \left(\pi^{-1/4} s^{-1/2} \right) e^{i \omega_0 \eta} e^{-\eta^2/2} \;,
\end{equation} where $\eta \equiv (t' - t)/s$ for scale $s$ and offset $t$ and $\omega_0$ chosen to equate the wavelet scale with its Fourier period $4 \pi s / \lambda = 4 \pi = \omega_0 + \sqrt{2+\omega_0^2}$, with a mother wavelet of unity scale and sample frequency, $\Psi_{1,0}(t') = \pi^{-1/4} e^{i \omega_0 t'} e^{-t'^2/2}$ for $\Delta t' \equiv 1$, and support with of $\pm \chi$ time units for $\chi=6$.  The amplitude of the continuous wavelet transform at a given scale and offset is then given by the weighted convolution with the signal $f$, \begin{equation}
{\rm CWT}(s,t) = \sum_{t'} s^{-1} \Psi_{s,t}^*(t') f(t') \;,
\end{equation} bringing the factor of $s^{-1}$ over from the inverse transform ({\it cf} Equations (6) and (9) in~\cite{Frick:1997426}), and we have found that the inclusion of a factor of 2 yields the power spectral density ${\rm PSD}(s,t) = |2 \times {\rm CWT}(s,t)|^2$ and will absorb that factor into the forward transform, ${\rm CWT} \rightarrow 2 \times {\rm CWT}$.  The inverse wavelet transform remains \begin{equation}
{\rm IWT}(t) = Re \left[\sum_s \sum_{t'} s^{-1} \Psi_{s,t}^*(t') {\rm CWT}(s,t') \Delta s \right] \;.
\end{equation}  Alternatively, one could introduce a factor $\sqrt{2}$ on the wavelet amplitude so as to keep the forward and inverse transforms symmetric, with another factor $\sqrt{2}$ applied to the transform amplitude when constructing the power spectral density.  The cone-of-influence is given by the $e$-folding time, $t_e = s \sqrt{2}$ for the Morlet wavelet, and indicates that region of the CWT where the amplitudes have become significantly affected by the truncation of the known signal.  We also define the cone-of-admissibility to be that region of the CWT whose wavelets are wholly contained by the duration of the signal $t \in \{1,\ldots,N_t\}$ and note that it is dependent upon the support width $\chi$.  For wavelets outside the cone-of-admissibility, we apply the edge correction algorithm \begin{equation}
{\rm CWT}(s,t) \rightarrow {\rm CWT}(s,t) \left[ \frac{\sum_{t'} \Phi_{s,0}(t')}{\sum_{t'} \Phi_{s,\tau}(t')} \right]^{1/2} \;,
\end{equation} for $t = 1 + \chi s - \tau$ along the left edge and $t = N_t - \chi s + \tau$ along the right, with a similar algorithm for wavelets exceeding the signal duration.  Unlike adaptive wavelets, \cite{Frick:1997426}, and the lifting scheme of \cite{sweldens98lifting}, this technique, similar to the weighted wavelet transform of \cite{Foster:1996}, preserves the shape (if not the moments) of the analyzing wavelet regardless of the distance to the edge, an important ingredient for the enhancement to follow.  

\begin{figure}
\noindent\includegraphics[scale=.9]{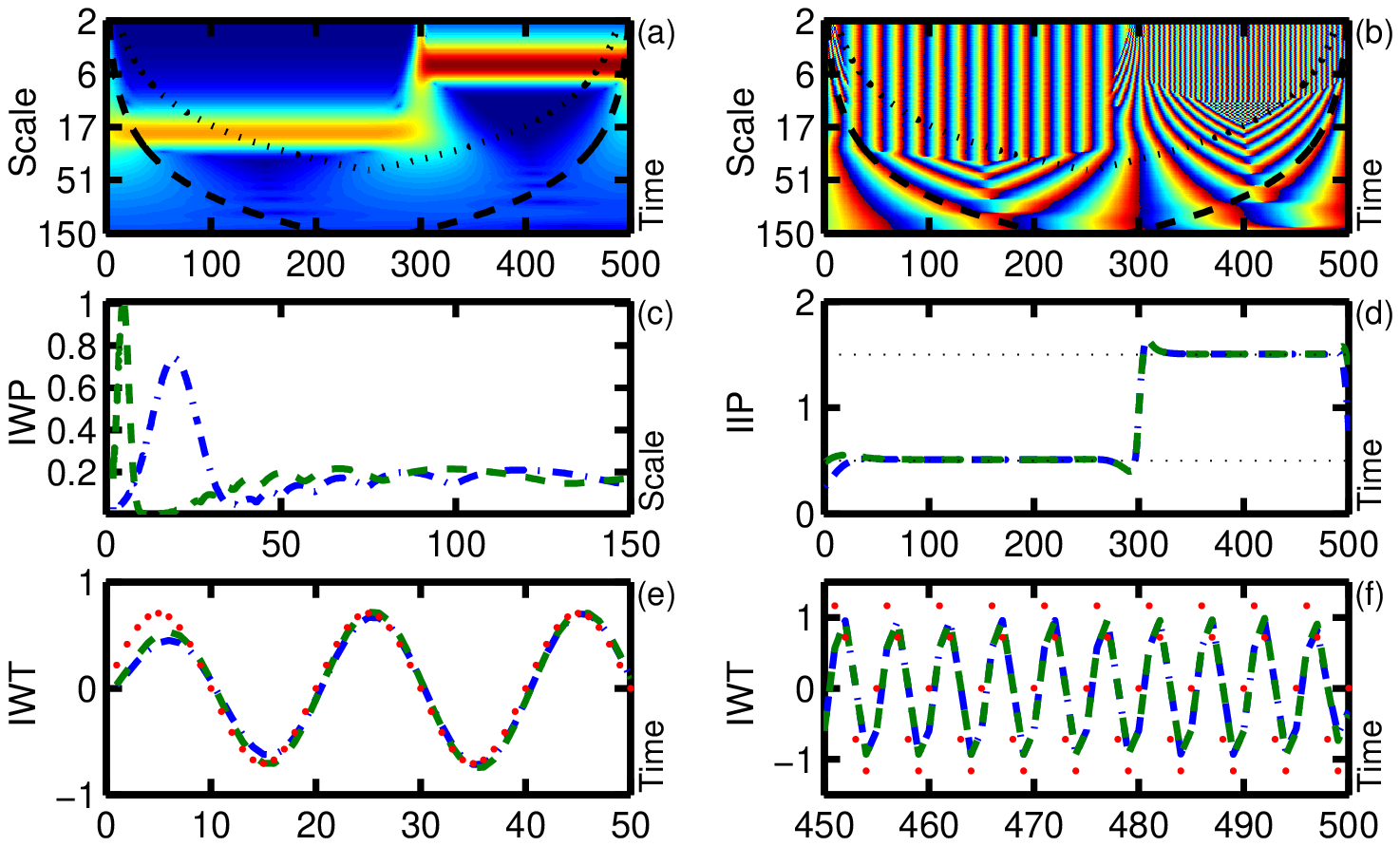}
\caption{\label{figA} (Color online.) Continuous wavelet transform analysis of a test signal of two finite sinusoids. (a) The magnitude of the CWT reveals each signal elements duration. The cone-of-influence and the cone-of-admissibility are given by the dashed and dotted lines. (b) The phase of the CWT aligns with the lines for the cones beyond the scale of excitation. (c) The instant wavelet power at times 150 (dash-dot) and 400 (dashed), displayed such that ${\rm IWP} = y^8$ so that the lobing at large scales becomes apparent. (d) The integrated instant power for the corrected (dashed) and uncorrected (dash-dot) CWT agree with the square of the signal element amplitudes upto contributions from the lobing and edge effects. (e) and (f) The inverse wavelet transforms for the corrected (dashed) and uncorrected (dash-dot) CWT along with the test signal (dots).  The IWT reconstructs lower frequencies well beyond the first cycle and encounters difficulties for higher frequencies. }
\end{figure}

We first analyze a test signal of duration 500 units comprised of a period 20, amplitude $\sqrt{0.5}$ sine wave for 300 units followed by a period 5, amplitude $\sqrt{1.5}$ sine wave for 200 units.  In Figure~\ref{figA}(a) we show the magnitude of the uncorrected CWT, and in Figure~\ref{figA}(b) we display its phase.  The cone-of-influence and more restrictive cone-of-admissibility are given by the dashed and dotted lines overlaid.  The spread in power at the location of signal transition is similar to that induced by the edges of the data but is the result of the truncation of either signal element rather than of the sample record.  The instant wavelet power is simply ${\rm IWP}_t(s) = {\rm PSD}(s,t)$, and in Figure~\ref{figA}(c) for times at the midpoint of each signal element's duration we show its one-eighth power $[{\rm IWP}_t(s)]^{1/8}$ so as to bring the low frequency (large scale) lobing into view.  Figure~\ref{figA}(d) displays the integrated instant power, ${\rm IIP}(t) = \sum_s {\rm PSD}(s,t) \Delta s$, for which we find agreement with the square of the signal element amplitude upto the effects of the signal truncation.  The uncorrected transform's IIP falls of steeply at the edges of the data, while that of the corrected transform displays a slight increase.  The reconstruction of the signal along the edges is shown in Figure~\ref{figA}(e) and (f), where we find good reconstruction beyond the first cycle for low frequencies and a diminished reconstruction for higher frequencies beyond a scale greater than about 6.  The loss of reconstruction is attributed to the truncation of wavelet coefficients beyond the scale of the Nyquist frequency.

\begin{figure}
\noindent\includegraphics[scale=.9]{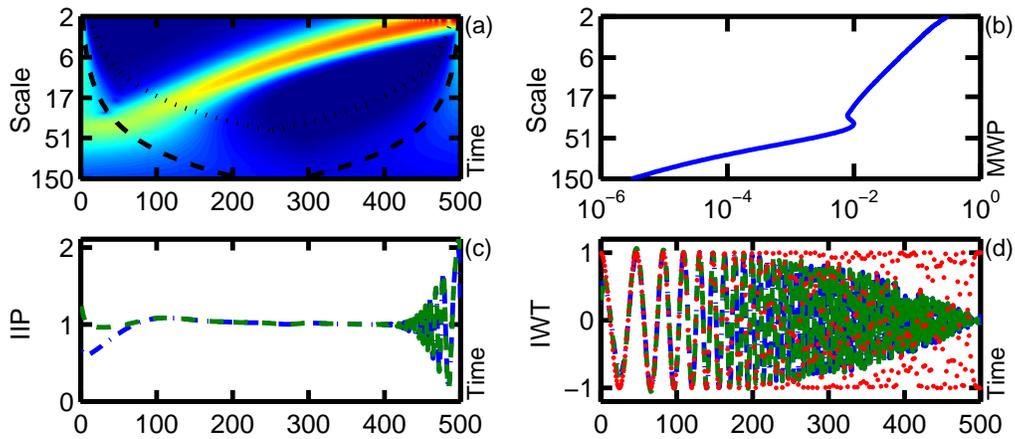}
\caption{\label{figK} (Color online.) Continuous wavelet transform analysis of a chirp test signal. (a) The magnitude of the CWT with the cone-of-influence and the cone-of-admissibility given by the dashed and dotted lines. (b) The mean wavelet power MWP. (c) The integrated instant power for the corrected (dashed) and uncorrected (dash-dot) CWT.  The IIP becomes oscillatory when the signal approaches the scale of the Nyquist frequency. (d) The inverse wavelet transforms for the corrected (dashed) and uncorrected (dash-dot) CWT along with the test signal (dots).  The loss of reconstruction amplitude is attributed to the truncation of the wavelet coefficients beyond the scale of the Nyquist frequency.}
\end{figure}

We next analyze a test signal of duration 500 units comprised of an oscillation with frequency increasing with time from period 50 to period 2.  In Figure~\ref{figK}(a) we show the magnitude of the corrected CWT, and in Figure~\ref{figK}(b) we give the mean wavelet power, ${\rm MWP}(s) = N_t^{-1} \sum_t {\rm PSD}(s,t)$.  Figure~\ref{figK}(c) displays the IIP for the corrected and uncorrected transforms, which displays oscillation around the square of the amplitude as the test signal approaches the Nyquist frequency.  The inverse transforms shown in Figure~\ref{figK}(d) display a loss of reconstruction amplitude as the extent of the transform response surpasses the scale of the Nyquist frequency.  What we have learned is that the continuous wavelet transform works best when the signal to analyze is formally periodic and known to be bandwidth limited to that region well resolved by the transform spectrum.  Nonetheless, we will apply the transform to finite signals of unknown bandwidth as a matter of practicality.

\section{Enhanced Wavelet Transform}
\begin{figure}
\noindent\includegraphics[scale=1]{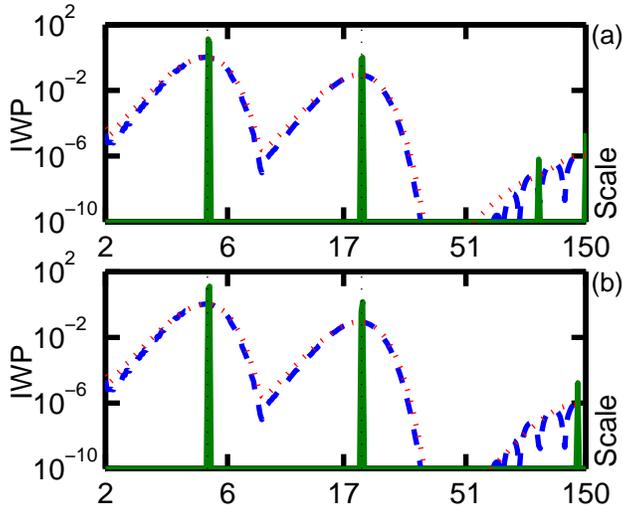}
\caption{\label{figB} (Color online.) Instant wavelet power (dashed), enhanced instant power (spikes), and instant power response (dotted) at the midpoint of a test signal of two sinusoids. The signal peaks are well resolved, and the low frequency lobing is not well modeled by the response of an instant signal. (a) Unit prior $P=1$. (b) Prior $P=1/(s^2 \Delta s)$. }
\end{figure}

One may treat the power spectral density coming from the CWT as the response of a theoretical apparatus---given a signal with a Dirac distribution in Fourier space, its representation in wavelet space is as a peak with integrated area equal to the power of the signal---thus opening the door to the techniques of resolution enhancement common in the analysis of experimental data; \cite{Sivia:1996} gives an excellent introduction.  The point spread function given by the wavelet response matrix $R(s,s')$ is defined as the integrated power of a wavelet of scale $s$ convoluted with a signal element of period $\lambda = s'$.  Here (note the factor of 2 mentioned above), we take the signal elements to be Hann windowed cosine functions for the duration of the wavelet, \begin{equation}
R(s,s') = |2 s^{-1} \sum_{t'} \Psi^*_{s,0}(t') \cos(2 \pi t' / s') H_s(t')|^2 \;.
\end{equation}  The enhanced instant power, then, is the solution to the equation \begin{equation}
0 = \sum_s P(s) \left[ \sum_{s'} R(s,s') {\rm EIP}_t(s') \Delta s' - {\rm IWP}_t(s) \right]^2 \Delta s \;,
\end{equation} where the prior $P(s)$ represents a scale-dependent weighting, found in a least-squares sense with non-negativity constraints.  There is nothing magical about the selection of the prior, which may be adjusted to weight various parts of the spectrum during the solution; however, one must be wary of results which depend significantly on that selection.  The enhanced wavelet transform is simply the collection of enhanced instant powers, ${\rm EWT}(s,t) = \sqrt{{\rm EIP}_t(s)}$.  Note that the redundancy of the CWT provides the resolution of the EWT, and that use of an alternate edge correction scheme would necessitate an offset dependent response matrix $R(t,s,s')$.  In Figure~\ref{figB}, we consider a test signal of 500 units comprised of the two sinusoids first mentioned for its duration, and we select the IWP at the midpoint to enhance.  The dashed line is the IWP; the spikes are the EIP; and the dotted line is the response power of the EIP.  For Figure~\ref{figB}(a) we select a unit prior, and for Figure~\ref{figB}(b) we use $P(s) = 1/(s^2 \Delta s)$.  We see that the signal peaks are well resolved, with a barely discernable shift for the higher frequency, and that the low frequency lobing is not well modeled by the response of an instant signal.

\section{Analysis of the Solar Magnetic Record}
\begin{figure}
\noindent\includegraphics[scale=1]{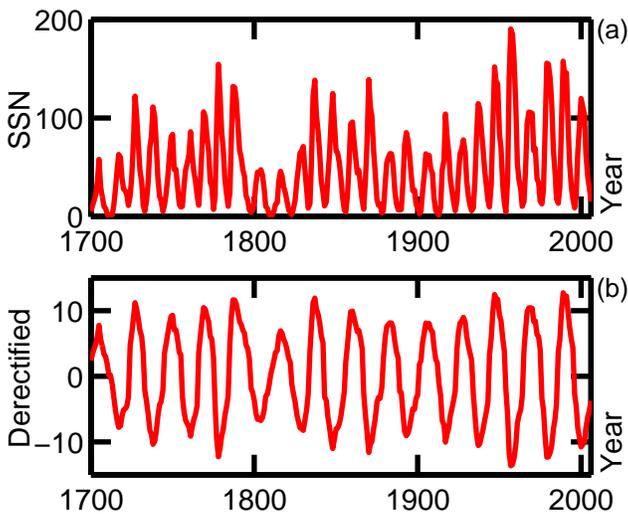}
\caption{\label{figC} (Color online.) (a) Yearly Smoothed Sunspot Number given by the Solar Influences Data Analysis Center,~\cite{sidc:2006}. (b) Derectified signal of solar magnetic activity. }
\end{figure}

\begin{figure}
\noindent\includegraphics[scale=.9]{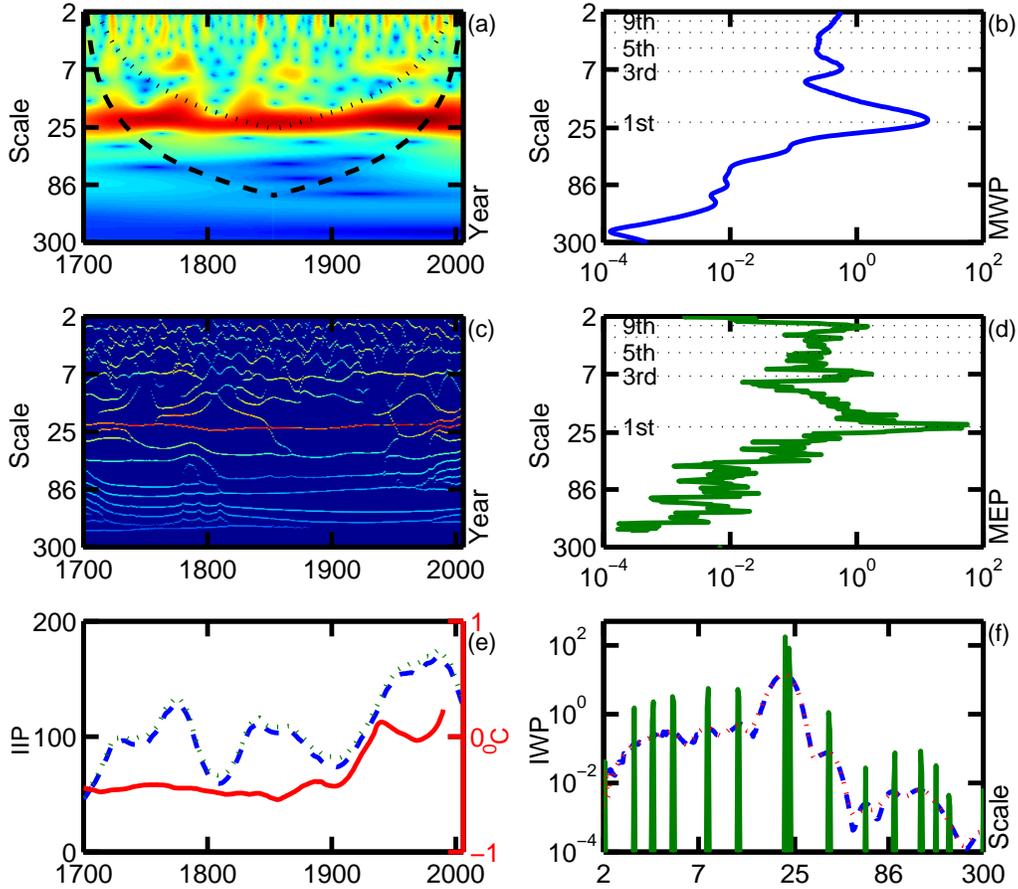}
\caption{\label{figD} (Color online.) Analysis of solar magnetic activity. (a) Power spectral density CWT, with the cones of influence and admissibility overlaid. (b) Mean wavelet power of the CWT, revealing the presence of the third harmonic of the 22 year Hale cycle. (c) Power spectral density EWT, revealing the interplay among harmonic cycles. (d) Mean wavelet power of the EWT, displaying prominent peaks at the third and ninth harmonics. (e) Integrated instant power for the CWT (dashed) and EWT (dotted) alongside the glacial reconstruction global temperature by~\cite{Oerlemans:2005}. (f) Instant wavelet power (dashed), enhanced instant power (spikes), and instant power response (dotted) at the year 1850.  Broad peaks in the CWT have been replaced by spikes in the EWT, giving a best estimate of the period of the signal elements. }
\end{figure}

For our signal of solar magnetic activity, we analyze the yearly record of sunspots given by the Smoothed Sunspot Number, SSN, at the Solar Influences Data Analysis Center \cite{sidc:2006}, reproduced in Figure~\ref{figC}(a).  Following~\cite{BuckandMac:1993} and~\cite{bracewell-53}, we derectify the signal by taking the square root of the SSN and inserting alternating signs as appropriate, thus revealing the Hale cycle of about 22 years, and subtract the mean before entering the wavelet analysis, shown in Figure~\ref{figC}(b).  The power spectral density of the CWT is displayed in Figure~\ref{figD}(a), where the fundamental cycle is readily apparent.  The mean wavelet power shown in Figure~\ref{figD}(b) reveals the presence of the third harmonic, while activity at higher frequency visible in the CWT has been blurred by averaging---dotted lines at the odd harmonics of a 22 year cycle are provided to guide the eye.  The EWT using a unit prior, displayed in Figure~\ref{figD}(c), reveals the yearly variation in period of the signal elements, and its mean enhanced power (MEP) in Figure~\ref{figD}(d) indicates that the ninth harmonic, while intermittent, carries nearly as much power as the third, while the seventh and fifth harmonics are suppressed.  Figure~\ref{figD}(e) presents the integrated instant power for the CWT and EWT as the dashed and dotted lines, respectively, and for amusement alongside we plot the glacial reconstruction of global temperature by~\cite{Oerlemans:2005}.  We observe that the IIP is dominated by the contribution from the fundamental, in agreement with results presented by~\cite{cjaa:35391} and~\cite{cjaa:46578}.  The IWP (dashed) and EIP (spikes) of Figure~\ref{figD}(f) are taken at year 1850, and we believe the large scale excitations represent an actual signal, as the response (dotted) fairly well represents the IWP at those scales, while admitting that the robustness of the spectrum for alternate priors remains to be investigated.  Comparison to other wavelet analyses of the sunspot record, such as by~\cite{Frick:1997670F}, is made complicated by the usual lack of derectification as well as the variety of transform normalizations used; however, we note the agreement with~\cite{BuckandMac:1993} on the presence of the third harmonic in the solar magnetic activity spectrum.

\begin{figure}
\noindent\includegraphics[scale=.9]{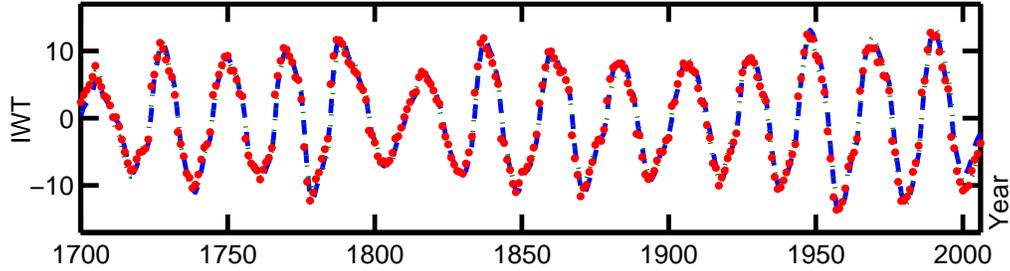}
\caption{\label{figE} (Color online.) Reconstruction for the inverse wavelet transform of the magnetic activity CWT (dashed) and EWT (dotted) as well as the original signal (dots). Both transforms fairly well represent the original signal.}
\end{figure}

Reconstruction from the EWT is so far defined upto an overall normalization using the original Morlet wavelet basis.  The inverse enhanced transform makes use of the magnitude of the EWT and the complex phase of the CWT, \begin{equation}
{\rm IET}(t) \propto \sum_s Re \left[ \sum_{t'} s^{-1} \Psi^*_{s,t}(t') {\rm EWT}(s,t') e^{i Ph[{\rm CWT}(s,t')]} \right] \Delta s \;,
\end{equation} where the constant of proportionality is chosen to minimize the discrepancy with the original signal.  In Figure~\ref{figE} we display the IWT (dashed) and IET (dotted) along with the original signal (solid).  The IWT is diminished slightly for the first and last half cycles of the fundamental, while the IET follows more closely the original signal; both reconstructions provide an accurate representation of the data signal over the majority of the time span.

\begin{figure}
\noindent\includegraphics[scale=1]{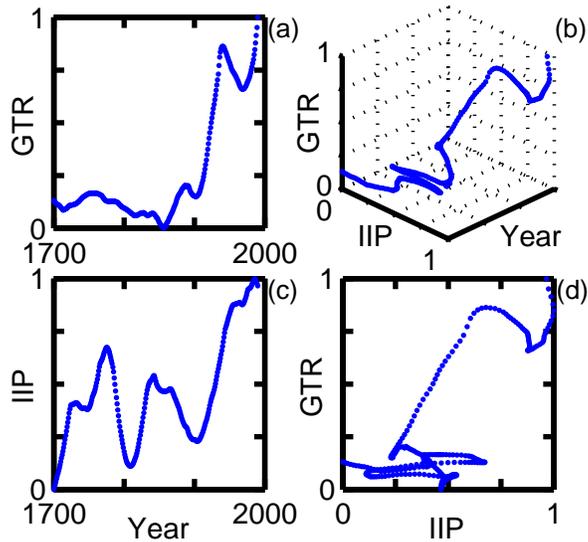}
\caption{\label{figF} (Color online.) Scatter plot of the unit normalized global temperature reconstruction (GTR) and integrated instant power (IIP) for the years 1700 to 1990. (a) and (c) Constituent quantities. (b) View from off-axis. (d) Global temperature {\it versus} integrated instant power. No correlation is apparent for the first two centuries, followed by a century of rough correlation.  }
\end{figure}

To compare the integrated instant power with the global temperature reconstruction for the years 1700 to 1990, we first normalize each to unit range then plot them both as a function of time, Figure~\ref{figF}.  Directing one's view along the time axis gives the scatter plot shown in Figure~\ref{figF}(d).  For the first two centuries, no dependence of global temperature on instant power is observed, followed by a sharp increase in both for the last century.  Whether the atmospheric changes following the Industrial Revolution are directly responsible for the increase in temperature or have made the climate more susceptible to influence from the solar magnetic activity as suggested by~\cite{msait:761026} is not revealed by this diagram, but we note that the solar activity has recently been at its highest level for three centuries, or longer if one includes the Maunder minimum of the century preceding this analysis.

\section{Comparison to the Central England Temperature Record}
\begin{figure}
\noindent\includegraphics[scale=1]{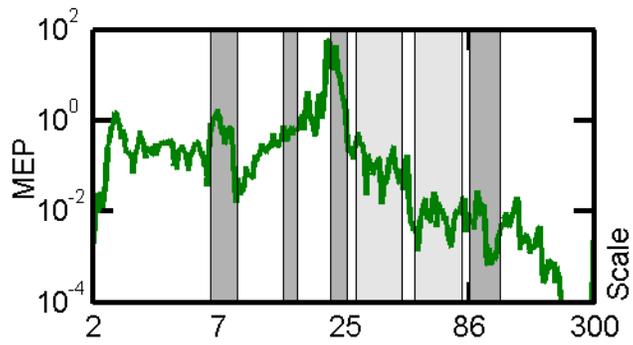}
\caption{\label{figG} (Color online.) Mean enhanced power of solar magnetic activity with shaded regions corresponding to the peaks found by~\cite{Baliunas:grl2411} in the CET.  The darker shading indicates more significant peaks.}
\end{figure}

\begin{figure}
\noindent\includegraphics[scale=.9]{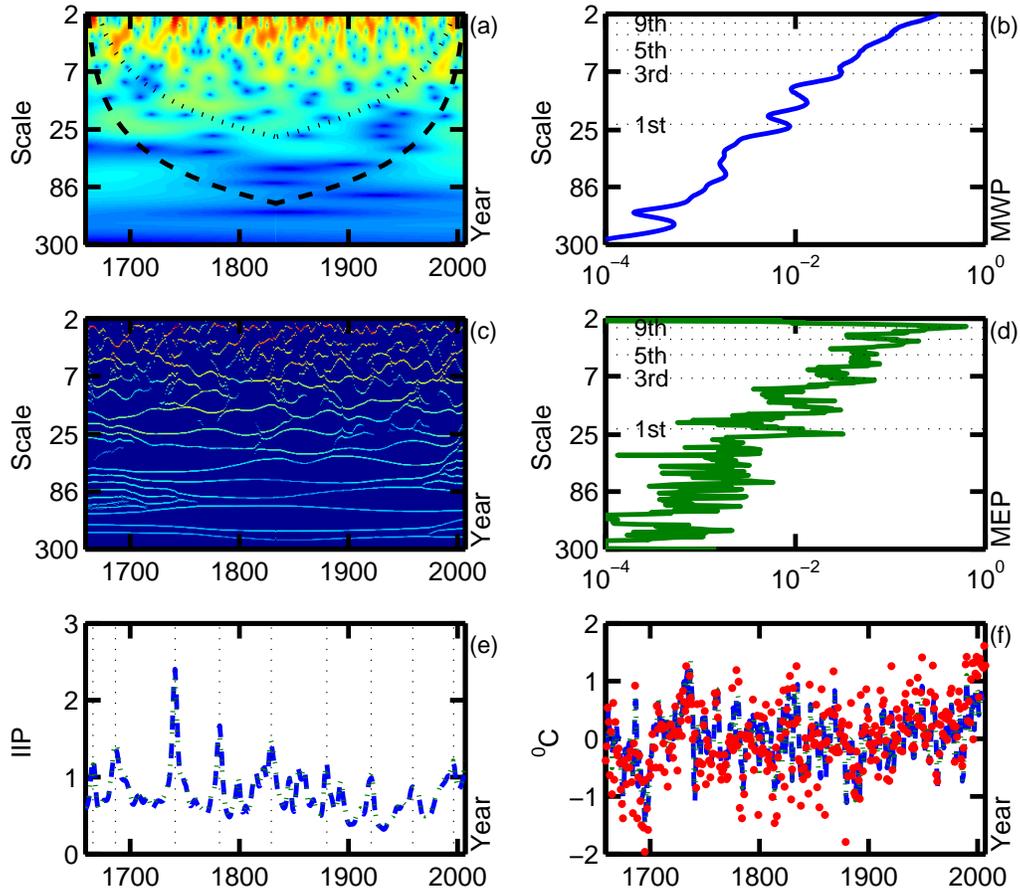}
\caption{\label{figH} (Color online.) Analysis of the Central England Temperature record for the years 1659 through 2007. (a) Power spectral density CWT, with the cones of influence and admissibility overlaid. (b) Mean wavelet power of the CWT with the odd harmonics of the 22 year Hale cycle overlaid. (c) Power spectral density EWT. (d) Mean enhanced power MEP indicating presence of several odd harmonics of the Hale cycle in the CET. (e) Integrated instant power of the temperature fluctuations, with indicated peaks approximately 44 years apart. (f) Reconstruction for the IWT (dashed) and IET (dotted) as well as the original signal (dots).  The poor reconstruction is attributed to the significant high-frequency content of the data.}
\end{figure}

Next we compare the mean enhanced power of the solar magnetic activity to the results of the analysis by~\cite{Baliunas:grl2411} of the Central England Temperature (CET) record, where significant peaks are found at yearly scales of $7.5\pm1$, $14.4\pm1$, $23.5\pm2$, and $102\pm15$, and others at $36\pm8$ and $65\pm15$, as shown in Figure~\ref{figG}.  The significant peaks other than at 14.4 years have corresponding analogues in the solar activity spectrum, suggesting a direct comparison of the enhanced wavelet spectra is in order.  Using the CET for years 1659 to 2007 by~\cite{hadcet:74} and~\cite{hadcet:92}, in Figure~\ref{figH}(a-d) we display its CWT, MWP, EWT, and MEP, respectively, where the odd harmonics of the Hale cycle are again shown by thin, dotted lines.  The correspondence of peaks at the fundamental, third, seventh, and ninth harmonics of the Hale cycle is suggestive of an interaction between solar magnetic activity and Earth's temperature.  The IIP of the temperature fluctuations is shown in Figure~\ref{figH}(e), where the indicated peaks occur roughly ever 44 years, and in Figure~\ref{figH}(f) are the IWT and IET alongside the data values.  We note that large amplitude fluctuations apparent in the data record are not well-represented by the sinusoidal signal elements of the inverse wavelet transform.

\begin{figure}
\noindent\includegraphics[scale=1]{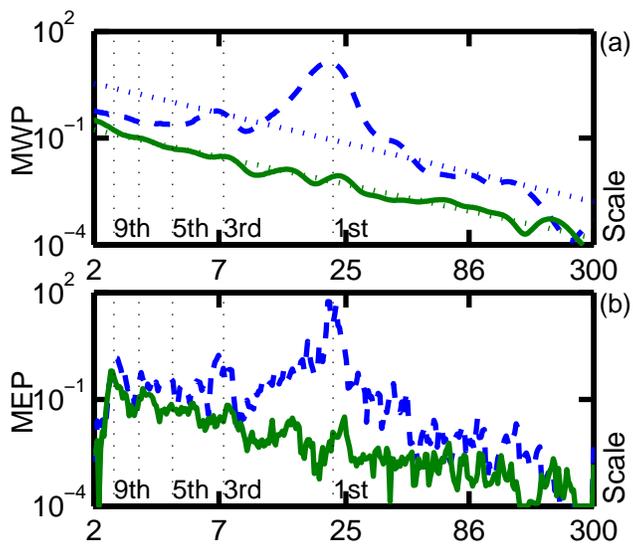}
\caption{\label{figI} (Color online.) Spectra for the solar activity and the CET compared to the odd harmonics of the Hale cycle. (a) Mean wavelet power for the solar activity (dashed) and the CET (solid) along with lines of best fit (dotted) giving power laws of -1.53 and -1.40 respectively. (b) Mean enhanced power for the solar activity (dashed) and the CET (solid).  The coincidence of peaks at the harmonics of the Hale cycle may or may not be significant.}
\end{figure}

\begin{figure}
\noindent\includegraphics[scale=1]{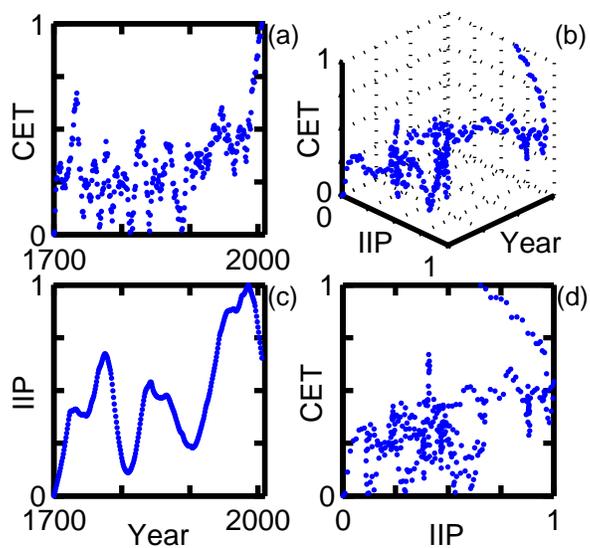}
\caption{\label{figJ} (Color online.) Scatter plot of the unit normalized 9-year running averaged CET and the solar activity IIP for the years 1700 to 2006. (a) and (c) Constituent quantities. (b) View from off-axis. (d) CET {\it versus} integrated instant power.  Whether any correlation exists is a matter for debate.}
\end{figure}

A direct comparison of the mean spectra for the solar magnetic activity with that of the CET is given by Figure~\ref{figI}.  Along with the continuous MWP in Figure~\ref{figI}(a) we show the corresponding power laws with exponents of -1.53 for the solar activity and -1.40 for the CET, where our wavelet scaling results in a power law with exponent -2 for Gaussian noise.  Compared to the value found by~\cite{Baliunas:grl2411} of -0.6, a factor of -1 is attributable to the difference in scaling of the wavelet transforms used.  When displayed together in Figure~\ref{figI}(b), the corresponding MEP's show several similarities at the odd harmonics of the Hale cycle; however, the peaks do not appear at precisely the same scales, leaving suggestion of their correlation tentative.  For our last comparison, we plot the unit normalized 9-year running averaged CET against the normalized solar magnetic IIP as functions of time for the years 1700 through 2006 in Figure~\ref{figJ}.  The scatter plot of Figure~\ref{figJ}(d) again suggests a possible correlation for the last century depicted; however, the recent decline apparent in solar magnetic activity does not correlate with the recent increase in the temperature measurements.

\section{Conclusions}
Using the power spectral density constructed from the continuous wavelet transform, an enhanced spectral analysis reveals the strong presence of the third and ninth harmonics of the fundamental Hale cycle in the derectified yearly smoothed sunspot record as well as the intermittency of the signal elements.  The integrated instant power is compared to a record of global temperature, indicating no correlation before the 20th century followed by a possible correlation since.  Repeating the analysis on the Central England Temperature record indicates that the Hale cycle may be represented in its measurements.  How the intensity of solar magnetic activity influences global climate remains an open question.


%
%
%
%
%
%

%
%
%
%

\subsection*{Acknowledgments}
Sunspot data provided by the SIDC-team, World Data Center for the Sunspot Index, Royal Observatory of Belgium, Monthly Report on the International Sunspot Number, online catalogue of the sunspot index: http://www.sidc.be/sunspot-data/, 1700--2006.  Global temperature data provided by J. Oerlemans, 2005, Global Glacier Length Temperature Reconstruction, IGBP PAGES/World Data Center for Paleoclimatology, Data Contribution Series \#2005-059, NOAA/NCDC Paleoclimatology Program, Boulder CO, USA.  Central England Temperature record provided by the Met Office Hadley Centre for Climate Change, http://www.metoffice.gov.uk/hadobs.

%
%
%
%
%
%
%
%
%
%


%
%


\clearpage


\end{document}